\documentclass[useAMS,usenatbib]{mn2e}
\usepackage{graphicx}
\usepackage{amsmath,amssymb}
\newcommand{\kms}{\>{\rm  km}\,{\rm  s}^{-1}}

\newcommand{\hMpc}{{\>h^{-1}\rm  Mpc}}

\title[Clustering of HI galaxies]
      {Clustering of HI galaxies in HIPASS and ALFALFA}
      \author[S.~S. Passmoor et al.]{S.~S. Passmoor$^{1}$\thanks{E-mail:sean.passmoor@gmail.com}, C.~M.~Cress$^{1,2}$, A.~Faltenbacher$^{1}$ \\
$^{1}$Physics Department, University of the Western Cape , Bellville 7535, South Africa\\$^{2}$Centre for High Performance Computing, 15 Lower Hope St, Cape Town 7700, South Africa
}
\begin{document}
\maketitle
\begin{abstract}
  We investigate the clustering  of HI-selected galaxies in  the
  ALFALFA survey and compare results with those obtained for HIPASS.
  Measurements  of  the angular  correlation 
  function and  the inferred  3D-clustering are compared  with results
  from   direct  spatial-correlation   measurements. We are able to
  measure clustering on smaller angular scales and for galaxies with
  lower HI masses than was previously possible. We calculate the
  expected clustering of dark matter using the redshift distributions
  of HIPASS and ALFALFA and show that the ALFALFA sample is somewhat
  more anti-biased with respect to dark matter than the HIPASS sample. 
\end{abstract}
\begin{keywords}
  large-scale structure of the universe - radio lines: galaxies
\end{keywords}
\section{Introduction}
Measurements of the clustering of galaxies allows one to investigate
the relationship between dark and luminous matter. By comparing
galaxies selected in different ways one gains understanding of how
different galaxies trace the underlying dark matter and also of
processes at work in galaxy evolution. This information is important when
using galaxies as probes of cosmological parameters. 

A number of new radio telescopes, such as the
MeerKAT\footnote{www.ska.ac.za} ,
ASKAP\footnote{www.atnf.csiro.au/SKA/}  and the 
SKA, are in the pipeline and they will detect huge numbers of galaxies
using HI. A reliable measure of the bias of HI-selected galaxies and
insight into the evolution of the bias is important for forecasting
the capabilities of telescopes which will probe HI at intermediate or
high-redshifts. 
The clustering of HI-selected galaxies has been studied by
\cite{Meyer-07}, \cite{Basilakos-07} and \cite{Ryan-Weber-06}. They
used data from the HI Parkes All Sky Survey (HIPASS,
\citealt{Meyer-04}), a blind survey for HI of the southern sky which
generated a catalogue of 4315 sources, the bulk of which have
redshifts below $z\sim0.02$. They showed that HI-selected galaxies are
less clustered than galaxies selected in other ways. \cite{Meyer-07}
investigated clustering of various subsamples of HIPASS galaxies,
showing that galaxies with high rotation velocities are more clustered
than those with lower rotation velocities. There were indications that
galaxies containing more HI are also more clustered but the
differences were not as pronounced as in \cite{Basilakos-07}. The
latter work also measures the bias of HIPASS galaxies relative to the
expected dark matter distribution. 

In this paper we measure the clustering of HI-selected galaxies
detected with the Arecibo L-band Feed Array (ALFA) and compiled in the
partially completed ALFALFA survey (the Arecibo Legacy Fast ALFA survey,
\citealt{Giovanelli-05a}). The results are compared with those obtained
for HIPASS. Clustering measurements in HIPASS are limited to large
angular scales where the beam-size of $\sim 15$ arcmins does not cause
confusion. The ALFALFA resolution is more than four times better
allowing us to probe clustering on smaller scales. The rms noise per
ALFALFA beam is about six times smaller, providing a catalogue
of sources which spans a wider range of redshifts and includes
galaxies with lower HI masses. We are thus able to measure clustering
of HI-selected galaxies in regimes that have not yet been explored and
to investigate trends seen in HIPASS, using an independent survey.  

The outline of the paper is as follows: In \S~\ref{sec:Data} we give a
short introduction to the HIPASS and ALFALFA surveys.  The computation
of the  angular and spatial two-point  correlation functions is described
in \S~\ref{sec:TwoPoint}.  The results are presented, discussed and
compared with earlier work in \S~\ref{sec:Results}. Finally,
\S~\ref{sec:Conclusions}  concludes with a short summary.
\section{Data}
\label{sec:Data}
HIPASS covers all the southern sky with  $\delta < +\,2^\circ$ and and can detect HI with velocities in the range $300\kms  - 12700\kms$.   The  rms noise  per beam  is
$\sim13$ mJy  \citep{Meyer-04}.  To exclude  structure associated with
the Milky-Way,  like high-velocity clouds  and low mass satellites, we
only use sources with recessional velocities larger than $600\kms$.
The average mass of HI in HIPASS galaxies is $3.24\times10^9 \rm M_\odot$.

When completed,  the ALFALFA Survey  \citep{Giovanelli-05a} will cover
$7000\deg^2$  of sky with  high galactic  latitude and  to a  depth of
$cz\sim18000\kms$. The rms noise of the survey is $\sim 2.2$ mJy and the beam-size is $\sim 3.6$ arcminutes. Currently the ALFALFA survey  contains three strips covering a total area of $\sim 400\deg^2$. These are the  two strips  centred on the  Virgo 
region and  the anti-Virgo strip.   
They contain  1796 sources  with $cz  > 600\kms$.   The first
completed Virgo  strip is defined by  $11^h44^m < \alpha  < 14^h00^m $
and $12^{\circ}  < \delta  < 16^{\circ} $ and contains $708$ sources.  The second Virgo strip contains $556$ galaxies within $11^h36^m < \alpha
< 13^h52^m $ and $8^{\circ} < \delta < 12^{\circ}$ \citep{Kent-08}.
The  anti-Virgo strip contains $488$ sources within $22^h00^m < \alpha  < 03^h04^m $ and
$26^{\circ} <  \delta < 28^{\circ}$  \citep{Saintonge-08}.  A circular
region, with radius of $1^{\circ}$ centred on M~87, has been removed from
the survey area due to the interference of M~87 \citep{Giovanelli-07}.
The average ALFALFA HI mass is $2.48\times10^9 \rm M_\odot$. The source
      density of the ALFALFA catalogue is approximately 20 times higher than that
      of the HIPASS survey.  
\begin{figure}
  \begin{center}
    \includegraphics[scale=0.6]{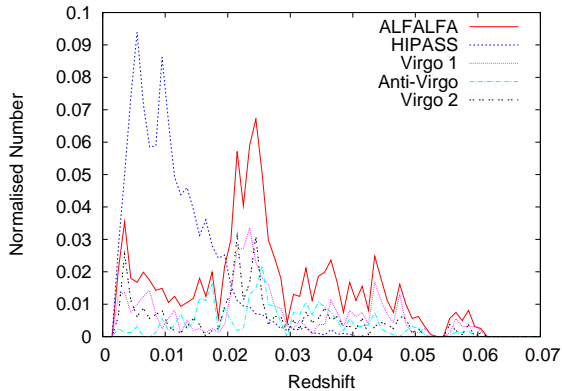} 
    \caption{\label{fig:norm} Plot  of the normalised redshift distribution of
      the HIPASS and ALFALFA
      surveys as well as  the distribution  of the  ALFALFA sources in the three,  spatially separated strips.}
  \end{center}
\end{figure}

Figure~\ref{fig:norm} displays
the  normalised  redshift distributions  of  the  HIPASS and ALFALFA  surveys as well as the redshift distributions in the three separate ALFALFA regions.   A redshift  of $z=0.02$  corresponds  to $\sim70\hMpc$
which  is roughly  the distance  to the  Coma supercluster. The high galaxy density near the Coma cluster and Virgo cluster are evident in the redshift distributions shown for the Virgo regions.  In the
anti-Virgo region the effect of the Perseus-Pisces supercluster can also be
seen as a slight enhancement of galaxies at a redshift of about $z \approx
0.025$.  %Thus the  peaks in the redshift distributions  in the ALFALFA
%survey at $z\approx 0.025$ result  from the high galaxy density in the
%vicinity of the Coma and the Perseus-Pisces superclusters.  
Below, we
will discuss  the impact of these inhomogeneities  on the determination
of  the two-point  correlation  function. 
%The  shape  of the  HIPASS
%redshift distribution is determined by the sensitivity limit of the survey.
%
\section{Two-Point correlation functions}
\label{sec:TwoPoint}
Here we review some  basic properties of  the angular and
projected  two-point   correlation  functions ($\omega$  and  $\Xi$
respectively)  and indicate  their relations  to the  three dimensional (3-D)
real-space  two-point  correlation function, $\xi$.   Subsequently, we
introduce  the estimator used  here and  discuss the  construction  of the
random samples. We do not employ the weighted correlation functions
used in \cite{Meyer-07} as we are interested in comparing the results
of the two surveys and in comparing our results with those predicted
for dark matter within a $\Lambda$CDM model. The unweighted
measurements suffice for this work and we are able to check our
unweighted results against those of \cite{Meyer-07}.  
\subsection{The Angular Two-Point Correlation Function}
The angular correlation function, $\omega$, is a simple measure of the
clustering of galaxies as a function of angular separation on the sky,
$\theta$,  which  does  not   require  redshift  information.   It  is
calculated by counting galaxy  pairs within a given angular separation
bin and comparing this number to a corresponding figure derived from a
random  catalogue   with  the  same  area  and   shape.   The  angular
correlation,  $\omega(\theta)$,  gives the  excess
probability,  over random, of finding two galaxies separated by
angle $\theta$. 

If we assume a redshift-dependent power law describes the 3-D
real-space  correlation function,
$\xi(r,z)=(r/r_0)^{-\gamma}(1+z)^{\gamma-(3+\epsilon)}$ (as in 
\citealt{Peebles-80} and \citealt{Loan-Wall-Lahav-97})  then the angular
correlation function is related to the spatial correlation function by
the Limber equation  
\citep{Rubin-54,Limber-54}: 
\begin{equation}
  \left(\frac{\theta}{\theta_{0}}\right) ^{1-\gamma}=
  \frac{\int^{\infty}_{0}\frac{N^2(z)(1+z)^{\gamma-(3+\epsilon)}\sqrt{\pi} 
      (d(z)\theta)^{1-\gamma}}{d^\prime(z)r_{0}^{-\gamma}}
    \frac{\Gamma(-\frac{1}{2}+\frac{1}{2}\gamma)}
         {\Gamma(\frac{1}{2}\gamma)}}
       {\left( \int_{0}^{\infty}N(z)dz\right) ^2}\ ,\label{equ:limber}
\end{equation}
where  $d(z)$ is  the comoving  distance  and $N(z)$  is the  redshift
number     density     distribution     of    the     sources     (cf.
Figure~\ref{fig:norm}). We use $\epsilon  = 0.8$, consistent with the expected  clustering behaviour in linear  theory, although the surveys are so shallow that the evolution of $\xi$ could be ignored. The  measured values  for the
logarithmic  slope $a_\theta  = 1-\gamma$  and the  correlation length
$\theta_0$  ($\omega(\theta_0)=1$) can then be used to determine the 3-D parameters $r_0$ and $\gamma$.

The errors  for $\omega$  are calculated using  jack-knife re-sampling
\citep{Lupton-93}.   For this purpose the data  are split  up into  $N$ RA-bins  and the
correlation function is recalculated  repeatedly each time leaving out
a different bin. Thus a set of $N$ values 
$\{ \omega_i,i=1,...,N \}$ for the correlation  function are obtained  and the jack-knife error  of the
mean, $\sigma_{\omega_{mean}}$, is given by
\begin{equation}
  \sigma_{\omega_{mean}} = \sqrt{(N-1)\sum^N_{i= 1}(\omega_i-\omega)^2/N}\ .
\end{equation}
%
%$\{ m_{Ji},i=1,...,N \}$ for
%the correlation  function are obtained  and the standard error  of the
%mean, $\sigma_{J_{mean}}$, is given by
%
%\begin{equation}
%  \sigma^2_{J_{mean}} = (N-1)\sum^N_{i= 1}(m_{Ji}-m)^2/N\ .
%\end{equation}

The  HIPASS sample  has been  divided into  24 RA  bins while  for the
ALFALFA catalogue we use 12 bins such that each bin contains approximately
the same area of the sky.
\subsection{The Projected Two-Point Correlation Function}
The  projected correlation function,  $\Xi(\sigma)$, is  determined by
the number of  pairs at given radial and  projected separations, $\pi$
and  $\sigma$, and a subsequent  integration along  the
radial  direction.  For  that  purpose the  absolute  radial  distance
between  a pair  of  galaxies,  $\pi =  |(v_i-v_j)/H_0|  $, and  their
angular separation, $\theta$, are converted into a projected distance,
$ \sigma = [(v_i+v_j)/H_0]\tan(\theta/2)$. Thus,
\begin{equation}
  \frac{\Xi(\sigma)}{\sigma}
  =\frac{2}{\sigma}\int_{0}^{D_{limit}}\xi(\sigma,\pi)d\pi
\end{equation}
where $D_{limit}$ is the  limit where the integral converges. Here we set 
$D_{limit}  = 25\hMpc  \approx 2500\kms  $. The  projected correlation
function   is  related   to   the real-space   correlation  function   by
\citep[e.g.,][]{Davis-Peebles-83}:
\begin{equation}
  \frac{\Xi(\sigma)}{\sigma}
  =\frac{2}{\sigma}\int_{\sigma}^{\infty}\xi(r)\frac{rdr}{(r^2-\sigma^2)^{1/2}}\ .  
\end{equation}
Assuming that  the projected and real-space  correlation functions
follow   power   laws %$\Xi(\sigma)/\sigma=(\sigma/\sigma_0)^{-a_\sigma}$   and
%$\xi(r)=(r/r_0){^-\gamma}$, 
within the region of interest ($r < 10\hMpc$),
the parameter for  the real-space correlation function  can be derived
from the projected one by the following expression: 
\begin{equation}
  \frac{\Xi(\sigma)}{\sigma} 
  = \left(\frac{r_0}{\sigma}\right)^\gamma 
  \frac{ \Gamma \left( 1/2 \right) \Gamma \left( (\gamma-1)/2\right) }
  {\Gamma \left( \gamma/2 \right)}
\end{equation}
More specifically, we calculate $r_0$  and $\gamma$ by fitting a power
law,     $\Xi(\sigma)/\sigma=\left(\sigma/\sigma_0\right)^{-a_\sigma}$,    using
the Levenberg-Marquardt nonlinear  least-squares method. The
parameters of the real-space correlation function are then given by
%\beta\left(1/2,-(a+1)/2\right)\right]
\begin{equation}
  r_0 =  \sigma_0 \left[\frac{ \Gamma  \left( 1/2 \right)  \Gamma \left(
      (a_\sigma-1)/2\right)     }     {\Gamma     \left(     a_\sigma/2     \right)}
    \right]^{-\frac{1}{a_\sigma}}
\label{equ:r}
\end{equation}
\begin{equation}
    \gamma=a_{\sigma}.
\end{equation}
Therefore, similar  to the angular correlation  function the projected
correlation  function  can  be   used  to  determine  the  real-space
clustering.   We  apply  both  methods  to determine  the  real-space
clustering strength  based on the  HIPASS and the ALFALFA  surveys and
compare the results.
\subsection{Estimator and random sampling}
Three  different estimators  are commonly  used to  determine  the
two-point   correlation   function  \citep{Davis-Huchra-82,
  Hamilton-93, 
  Landy-Szalay-93}.   In this work  we use  the \cite{Landy-Szalay-93}
estimator  as it reduces  errors caused  by edges  and holes  within a
given catalogue. In particular this is important for the ALFALFA survey
with the hole caused by  M~87 \citep{Giovanelli-07} and the large edge
effects due to the three strips. The estimator is of the form:
\begin{equation}
  \label{equ:estimator}
  \xi(\textbf{r})=
  \frac{DD(\textbf{r})-2DR(\textbf{r})+RR(\textbf{r})}{RR(\textbf{r})}\ ,
\end{equation}
where  $\textbf{r}$ is  the  separation distance  which has  different
meanings  for the  different correlation  functions.  For  the angular
correlation  it  denotes  the  separation  angle,  $\theta$,  for  the
projected correlation function it is the projected distance, $\sigma$,
and  for  the real-space  correlation  it  indicates the  real-space
distance  $r$.  $DD(\textbf{r})$  is  the number  of data-data  pairs,
$DR(\textbf{r})$   is   the   number   of   data-random   pairs,   and
$RR(\textbf{r})$  is  the  number  of  random-random  pairs  all  with
separations $\textbf{r}$.

The random catalogues were  generated with uniform distributions on the
sky and redshift distributions which resemble the 
distribution  of recessional velocities  in the
survey   smoothed   using  kernel   density   estimation
\citep{Wand-Jones-95}.  Throughout this  work we use random samples that are
equal in size  compared to the corresponding data  set.  We repeat the
random  catalogue generation 20 times and  the calculation  of random  pairs in
order to reduce the variance from the random sampling.
\subsection{The angular correlation function of dark matter}
Based on Limber's equation, the redshift distributions of HIPASS and ALFALFA
and the expression for the non-linear power spectrum discussed in \cite{Peacock-Dodds-96}  
we predict the angular correlation function of dark
matter using the cosmological parameters given in \cite{Komatsu-09}.  
The bias parameter, $b$, at various angles is then determined by 
\begin{equation}
b=\sqrt{\frac{\omega_{HI}}{ \omega_{dark matter}}}\ . 
\end{equation}
\section{Results}
\label{sec:Results}
%
%In this section we compute angular and projected correlation functions
%for HIPASS  and ALFALFA. The ALFALFA  survey had a rms  noise of $\sim
%2.2$ mJy and a average detected HI mass of $2.48$x$10^9 \rm M_\odot $.
%The corresponding values for HIPASS are $\sim 13$ mJy and $3.24$x$10^9
%\rm  M_\odot  $ .   We  fit power  laws  to  the observed  correlation
%functions and  derive the real space clustering  strength expressed by
%the correlation  length, $r_0$,  and the logarithmic  slope, $\gamma$.
%In a subsequent section we  discuss the differences in the correlation
%functions  found  for the  three  spatially  separated  strips of  the
%ALFALFA survey.
%
\subsection{The full HIPASS and ALFALFA samples}
\subsubsection{Angular correlation functions}
\begin{figure}
  \begin{center}
    \includegraphics[width=0.85\hsize]{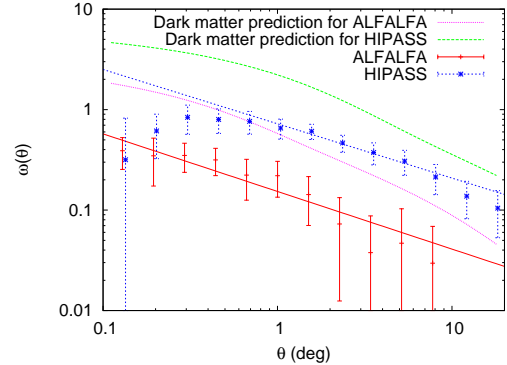} 
    \caption{\label{fig:angular}  Angular  correlation  functions  for
      HIPASS and ALFALFA.   Error bars
      were calculated using jack-knife  sampling. The solid red line and the small-dashed blue line
      show the corresponding power law fits for angles between $0.1$
      and $8^{\circ}$ for ALFALFA and between $1$ and $10^{\circ}$ for the HIPASS data. The projected clustering of dark matter (in a $\Lambda$CDM model) with redshift distributions of HIPASS and ALFALFA are shown by the green dashed line and the magenta dotted line respectively}
  \end{center}
\end{figure}
Figure~\ref{fig:angular} shows the angular
correlation functions for HIPASS and  ALFALFA data as well as the predicted correlation functions of cold dark matter weighted with the redshift distributions of the surveys.  The straight lines are power law
fits for pair  separations in the range between  $0.1$ and $8^{\circ}$ for
ALFALFA and  between $1$ and  $10^{\circ}$ for HIPASS. The effect of source confusion in HIPASS is evident at smaller angular scales.  The corresponding
parameters,     $\theta_0$     and     $a_{\theta}$,     are     given
in Table~\ref{tab:angular}.  
%Employing  Eq.~\ref{equ:limber} these values
%can  be translated into  real space  clustering parameters,  $r_0$ and
%$\gamma$ and are shown in Table~\ref{tab:realspace}.

%For  HIPASS and ALFALFA  we obtain, $r_0=2.89\pm0.08\hMpc$,
%$\gamma=1.56\pm0.02$ and $r_0=2.00\pm0.40\hMpc$, $\gamma=1.59\pm0.06$,
%respectively (see also Table~\ref{tab:realspace}).

The measured slopes in the two surveys agree reasonably well.  As
expected, the value  of  $\theta_0$ is lower for ALFALFA  since it is
deeper than HIPASS and the clustering in 3-D is washed out in the 2-D
projection. ALFALFA also detects galaxies with lower HI masses which
are potentially less clustered.  
% for a given angular separation
%it contains more distant  pairs of sources that reduce the amplitude of
%the angular correlation function compared to that derived from the HIPASS catal%ogue, in addition  
%the deeper flux limit probes galaxys with a lower average mass which cluster mo%re weakly. 
%
\begin{table}
  \begin{center}
    \begin{tabular}{ | l | l | l | }
      & HIPASS & ALFALFA \\\hline
      $\theta_0$& $0.603\pm 0.04^{\circ}$ & $0.044\pm 0.013^{\circ}$\\
      $a_{\theta}$  & $0.56\pm 0.02$     & $0.59\pm 0.06$
    \end{tabular}
  \end{center}
  \caption{\label{tab:angular} The angular clustering fitted parameters,
    $\theta_0$ and $a_{\theta}$.}
\end{table}

\begin{table}
  \begin{center}
    \begin{tabular}{ | l | l | l | }
      & HIPASS & ALFALFA \\\hline
      $\sigma_0$& $6.29\pm 0.36\hMpc$ & $5.34\pm 1.08\hMpc$\\
      $a_{\sigma}$  & $1.62\pm 0.04$  & $1.68\pm 0.13$
    \end{tabular}
  \end{center}
  \caption{\label{tab:projected} The projected clustering fitted parameters,
    $\sigma_0$ and $a_{\sigma}$.}
\end{table}

\subsubsection{Projected correlation functions}
\begin{figure}
  \begin{center}
    \includegraphics[width=0.85\hsize]{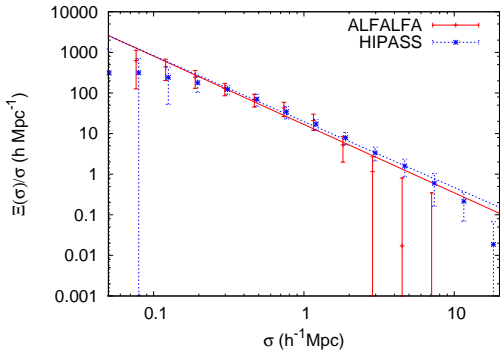} 
    \caption{\label{fig:projected} Projected correlation functions for
      HIPASS and ALFALFA.  Error bars
      are calculated  using jack-knife  sampling. Lines
      show the  corresponding power law fits for separations between
      $0.1$ and $3.5\hMpc$  for ALFALFA and $0.2$ and $8\hMpc$ for the HIPASS data.}
  \end{center}
\end{figure}
Figure~\ref{fig:projected}  shows the
projected correlation functions  for the  HIPASS and  ALFALFA surveys.
The  lines represent power law  fits and  the corresponding  parameters are
presented  in Table~\ref{tab:projected}.  Once again the slopes,
$a_\sigma$, agree well while the amplitude of clustering, $\sigma_0$,
in ALFALFA is lower (although the uncertainties are fairly large).    
%The agreement
%in the  clustering length,  $\sigma_0$ in both  surveys is due  to the
%fact that the projected distances are not affected by the depth of the
%survey.
%
\subsubsection{Inferred spatial correlations}
Table~\ref{tab:realspace} shows the spatial correlation function parameters inferred from the angular and projected correlation functions obtained using Eq.~\ref{equ:limber} and Eq.~\ref{equ:r}. The subscripts, $\theta$ and $\sigma$, indicate which correlation function has been used to derive these parameters. The two values obtained for $r_0$ in HIPASS are within 2$\sigma$ of each other and agree well with the unweighted value of 2.7 obtained by \cite{Meyer-04}. The two ALFALFA values are consistent with each-other and indicate somewhat lower clustering than HIPASS. 
\subsubsection{Bias estimation}
The predicted angular correlation function of dark matter is compared
with our results in Figure~\ref{fig:angular}. We have calculated the
bias for the two surveys at each data point in the plot. For HIPASS,
in the $1-10^{\circ}$ range, we find bias values ranging 
from 0.54 to 0.70, with an average of 0.63. This is fairly consistent
with the value of 0.68 obtained by \cite{Basilakos-07}. For ALFALFA,
on the same angular scales, the values range between 0.41 and 0.62,
with an average of 0.52. Our results thus indicate that the ALFALFA
sample is somewhat more anti-biased than the HIPASS sample. This is
consistent with the idea that ALFALFA includes galaxies with lower HI
mass which are less clustered than the higher mass galaxies detected
in HIPASS. We note, however, that the lower values are found at large
scales where the narrowness of the strips may effect measurements more
severely.  
%although the lower values are found at large scales where the narrowness of the strips may effect measurements more. This is consistent with the idea that ALFALFA includes galaxies with lower HI mass which are less clustered than the higher mass galaxies detected in HIPASS.
%
\begin{figure}
  \begin{center}
    \includegraphics[width=0.85\hsize]{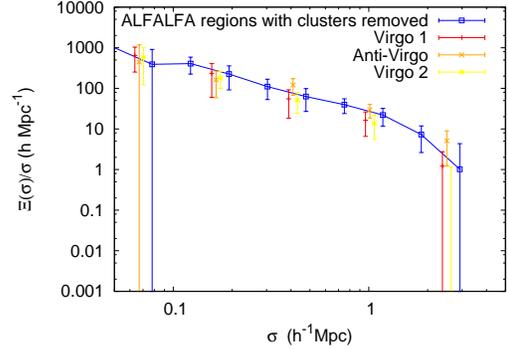}  
    \caption{\label{fig:comparison}   Comparison   of  the   projected
      correlation functions of the three strips of the current ALFALFA
      survey as well as the correlation function measured when ALFALFA data between $0.02<z<0.03$ is excluded (that is, when cluster galaxies in the Coma and Perseus-Pisces regions are excluded) 
      The data points for  the three strips are slightly offset
      to  improve   readability.  Error  bars   are  calculated  using
      jack-knife sampling.}
  \end{center}
\end{figure}
\subsection{Correlation functions of different ALFALFA subsamples} 
\subsubsection{Flux and HI mass subsets}
The ALFALFA  data has been subdivided into two equivalent parts  based
on the flux  of the sources.  For these subsamples the  angular and
projected correlation functions were recalculated  as  described in 
section  \ref{sec:TwoPoint}.  In agreement with  \cite{Meyer-07} 
we find that the two correlation  functions compare well with each
other and with the correlation function of the whole data set
indicating a negligible dependence of clustering on HI flux.

We also split the samples evenly into high and low HI-mass
subsamples. The clustering parameters obtained are shown in
Table~\ref{tab:realspace}. Our results for HIPASS are consistent with
those of \cite{Meyer-07}, indicating that the galaxies with higher
HI-masses are more clustered. Interestingly, the same trend is not
apparent in the ALFALFA survey but the uncertainties are fairly
large.% and we await further data to explore this further. 

We did not attempt to separate the galaxies according to their rotation
velocities as this requires additional data to estimate inclinations.  
\begin{table}
  \begin{center}
    \begin{tabular}{ | l | l | l | } % allthough the lines are commented out they have the updated value.
      &    HIPASS    &    ALFALFA    \\\hline\hline    $r_{0,\theta}$&
      $2.89\pm0.08\hMpc$  &  $2.00\pm0.40\hMpc$\\ 
      $\gamma_{\theta}$  & $1.56\pm0.02$    &   $1.59\pm0.06$   \\
\hline$r_{0,\sigma}$ & $2.51\pm0.20\hMpc$ &$  2.30\pm0.53\hMpc$ \\
      $  \gamma_{\sigma} $& $ 1.62\pm0.04$ & $ 1.68\pm0.13$ \\\hline
      &  $M_{HI} < 10^{9.25} {h^{-2} \rm M}_\odot$        \\\hline
      %$r_{0,\theta}$&$2.11\pm0.09\hMpc$  &  $2.43\pm0.26\hMpc$\\ 
      %$\gamma_{\theta}$  & $1.45\pm0.04$    &   $1.54\pm0.05$   \\
      $r_{0,\sigma}$ & $2.26\pm0.36\hMpc$ &$  2.48\pm0.69\hMpc$ \\
      $  \gamma_{\sigma} $& $ 1.60\pm0.08$ & $ 1.59\pm0.13$ \\\hline
      &  $M_{HI} > 10^{9.25} {h^{-2} \rm M}_\odot$ \\\hline
      %$r_{0,\theta}$ & $3.55\pm0.19\hMpc$  &  $2.52\pm0.52\hMpc$\\ 
      %$\gamma_{\theta}$  & $1.42\pm0.03$    &   $2.02\pm0.15$   \\
      $r_{0,\sigma}$ & $3.32\pm0.55\hMpc$ &$  2.04\pm0.65\hMpc$ \\
      $  \gamma_{\sigma} $& $ 1.50\pm0.08$ & $ 1.74\pm0.24$ \\
    \end{tabular}
  \end{center}
  \caption{\label{tab:realspace} The real-space clustering parameters,
    $r_0$ and  $\gamma$ derived from the  angular correlation function (indicated by subscript  $\theta$) and from  the projected
    correlation function (indicated by subscript $\sigma$.}
\end{table}
\subsubsection{Small Field effects}
%The ALFALFA data covers a fairly small region of sky and there is the
%concern that the clustering measured is not completely representative
%of the local volume. To investigate this further we compare the
%clustering in the three separate ALFALFA regions and show, in
%Figure~\ref{fig:comparison}, that the three separate measurements
%agree well, indicating that cosmic variance is not a big issue.  

The Virgo regions contain over-densities of galaxies that are associated with the Virgo and 
Coma clusters. There is the concern that the results will be 
biased by the presence of such dominant large-scale structure within
the relatively small survey fields. To investigate this, the
correlation functions  
of the three regions were calculated separately and are shown in Figure~\ref{fig:comparison}. 
Measurements in the three regions agree 
to within their uncertainties, indicating that 
the presence of the big clusters within the Virgo regions do not effect the results
significantly. 
%%This can be understood by considering that an increased density of
%%sources in the data field is balanced by an increased number of
%%random sources in the correlation function calculation. 
We note however, 
that the Anti-Virgo region is near the Perseus-Pisces supercluster
which causes a slight 
over-density in that field at a similar redshift ($z\approx0.025$). To be sure that over-densities in 
all three fields at this redshift were not biasing our results, we cut the galaxies with 
redshifts between $\sim0.02$ and $0.03$ out of the samples and recalculated the correlation 
functions. The results are also shown in Figure~\ref{fig:comparison}
and it is clear that the correlation functions   
with and without the redshift cuts are completely consistent within the uncertainties.

As an additional check on the effect of the small fields on the
measured clustering strength, we calculated the integral constraint
\citep{Peebles-80, Ratcliffe-98} for a single field, and obtained a
value of $0.145$ which indicates an effect within the uncertainty of
the correlation function derived from the ALFALFA data. 
\section{Summary and conclusions}
\label{sec:Conclusions}
We have measured the clustering of HI-selected galaxies using the
ALFALFA survey data and compared this with results for HIPASS. Our two
methods for determining the real-space correlation function agree well
and our results for HIPASS agree with those found by
\cite{Meyer-04}. The real-space clustering in ALFALFA appears to be
even lower than in HIPASS, consistent with the idea that ALFALFA
probes galaxies with lower HI-masses that are less clustered than
their high-mass counterparts. Our measurements of high- and low-mass
subsamples in ALFALFA do not provide evidence to support this idea but
the uncertainties on the measurements are large.

We have calculated the clustering of dark matter expected within a
$\Lambda$CDM model with redshift distributions of HIPASS and
ALFALFA. We then calculated the bias of ALFALFA sources over the range
$1-10^{\circ}$, finding a value of 0.62 at $1^{\circ}$ and an average
value of 0.52 over the whole range. The significant anti-bias of
galaxies with low HI-mass is important to consider when estimating the
signal-to-noise of experiments planned for the SKA and its
pathfinders.  

\section*{Acknowledgements}
We thank the South African Square Kilometre Array Project, National Research Foundation and Centre for High Performance Computing for support. We also thank the referee for helpful comments.  
%
%\bibliography{lit}

\label{lastpage}
%
%\newpage 
%\tableofcontents
%
%\newpage 
%\include{fig}
%
\end{document}